\documentclass[epj]{svjour}
\usepackage{times}
\usepackage{graphicx}
\usepackage{rotating}

\begin{document}

\title{Precise half-life measurement of the $^{26}$Si ground state}

\author{
I. Matea\inst{1} \and
J.~Souin\inst{1,2} \and
J.~\"{A}yst\"{o}\inst{3} \and
B. Blank\inst{1} \and
P.~Delahaye\inst{4} \and
V.-V.~Elomaa\inst{3} \and
T.~Eronen\inst{3} \and
J. Giovinazzo\inst{1} \and
U.~Hager\inst{3,}\thanks{Present address: TRIUMF, 4004 Wesbrook Mall, Vancouver, British Columbia, V6T 2A3, Canada} \and
J.~Hakala\inst{3} \and
J.~Huikari\inst{1} \and
A.~Jokinen\inst{3} \and
A.~Kankainen\inst{3} \and
I.D.~Moore\inst{3} \and
J.-L.~Pedroza\inst{1} \and
S.~Rahaman\inst{3} \and
J.~Rissanen\inst{3} \and
J.~Ronkainen\inst{3} \and
A.~Saastamoinen\inst{3} \and
T.~Sonoda\inst{3,}\thanks{Present address: Instituut voor Kern- en Stralingsfysica, Celestijnenlaan 200D, B-3001 Leuven, Belgium} \and
C.~Weber\inst{3}
}

\institute{
Centre d'\'Etudes Nucl\'eaires de Bordeaux Gradignan -
Universit\'e Bordeaux 1 - UMR 5797 CNRS/IN2P3, Chemin du Solarium,
BP 120, 33175 Gradignan Cedex, France
\and
Instituto Estructura de la Materia, CSIC, Serrano 113bis, E-28006 Madrid, Spain
\and
Department of Physics, University of Jyv\"askyl\"a, P.O. Box 35, FI-40014, Jyv\"askyl\"a, Finland
\and
CERN, CH-1211 Geneva 23, Switzerland
}

\abstract{
The ${\beta}$-decay half-life of $^{26}$Si was measured with a relative precision of 1.4$\cdot$10$^{-3}$. The measurement yields a value of 2.2283(27)~s which is in good agreement with previous measurements but has a precision that is better by a factor of 4. In the same experiment, we have also measured the non-analogue branching ratios and could determine the super-allowed one with a precision of 3\%. The experiment was done at the Accelerator Laboratory of the University of Jyv\"{a}skyl\"{a} where we used the IGISOL technique with the JYFLTRAP facility to separate pure samples of $^{26}$Si.
}


\maketitle

\section{Introduction}

Due to its inherent simplicity, the super-allowed nuclear $\beta$-decay between nuclear states with (J$^{\pi}$,T) = (0$^{+}$,1) is a
very powerful tool to test the present theory of weak interaction at low energies~\cite{hardy05}. This type of transition
depends to first order only on the vector part of the weak interaction. The corrected $\mathcal{F}t$ value, determined from
the experimental comparative life-time, \textit{ft}, is:

\begin{center}
\begin{eqnarray}\nonumber
  \mathcal{F}t &=& ft \times (1-\delta_C + \delta_{NS}) \times (1 + \delta'_R) \\
     &=& \frac{K}{g_V^2 \times \langle M_F \rangle^2 \times (1 + \Delta_R)}
\end{eqnarray}
\end{center}

and directly related to the vector coupling constant, $g$$_V$. The matrix element, $\langle$M$_F$$\rangle$, equals $\sqrt{2}$ for T=1 nuclei, \textit{ft} is determined from the mass difference between the initial and final analogue states, Q$_{EC}$, the half-life of the parent nucleus, T$_{1/2}$, and the branching ratio (BR) for the super-allowed decay, while $\delta_C$, $\delta_{NS}$, $\delta'_R$ and $\Delta_R$ are correction factors that must be determined by models~\cite{towner02,towner07}. \textit{K} is a constant.

From the corrected $\mathcal{F}t$ value, one can determine the vector coupling constant, $g$$_V$, and test the validity of the
Conserved Vector Current (CVC) hypothesis of the weak interaction stating that the vector part of the weak interaction is not
influenced by the strong interaction. Furthermore, the $g$$_V$ value combined with the weak vector coupling constant for the purely
leptonic $\mu$-decay, $g$$_V^\mu$, yields the up-quark down-quark element $V_{ud}$ of the Cabibbo - Kobayashi - Maskawa (CKM)
quark-mixing matrix:

\begin{equation}\label{Vud}
    V_{ud}^2 = \frac{g_V^2}{{g_V^\mu}^2} = \frac{K}{2 {g_V^\mu}^2 (1 + \Delta_R)\mathcal{F}t}
\end{equation}

Presently, this is the key ingredient in one of the most demanding tests of the unitarity of the CKM matrix that assures the validity of the three-generation Standard Model.

A recent review of super-allowed Fermi transitions reported such measurements in 20 nuclei with
(\overrightarrow{T};T$_z$) = (\overrightarrow{1};$-$1, 0)~\cite{hardy05}. Twelve nuclei have a precision close to or better
than $10^{-3}$ for the experimental ingredients needed and were used to determine $\mathcal{F}t$ with a precision close to $10^{-4}$. The reported average value is 3072.7$\pm$0.8~s \cite{hardy05}. This yields a $V_{ud}$ value of 0.9738(4). The nuclear $\beta$ decay provides the most precise determination of the up-quark down-quark element of the CKM matrix. We remind that $V_{ud}$ can also be determined from the neutron decay ($V_{ud}^{neutron}$ = 0.9746(18)) and from the pion beta decay ($V_{ud}^{\pi^{\pm}}$ = 0.9749(26)) \cite{PDG}.

Since the 2005 review of Hardy and Towner, the $^{62}$Ga super-allowed decay reached the required precision in order to increase to thirteen the number of transitions used to determine $\mathcal{F}t$ and its present adopted value is 3071.4(8)~s, leading to a value of 0.97418(26) for the $V_{ud}$ matrix element. These values incorporate also the most recent calculation for the correction factors \cite{towner07}.

What gives credit to the nuclear result for the g$_V$ value is the fact that a significant number of super-allowed transitions measured with high precision gives consistent results for $\mathcal{F}t$. An important work is in progress in order to add to the above mentioned 13 nuclei some of the other seven cited in~\cite{hardy05}. None of these seven nuclei has a precise measurement of the super-allowed BR due to the presence of Gamow-Teller transitions in competition with the super-allowed one. Concerning the half-life values, they are known with a relative precision worse than 2$\cdot$10$^{-3}$.

We report in this paper on half-life and BR measurements for the decay of the $^{26}$Si ($T_z$~= $-$1) nucleus. The aim of the experiment was to reach a precision of 10$^{-3}$ for the measured half-life.

Previous measurements of the $^{26}$Si half-life reported an average value of 2.234(12)~s~\cite{hardy75,wilson80}. The BR and the Q$_{EC}$ of the super-allowed decay are, respectively, 75.09(92)~\% and \\ 4836.9(30)~keV \cite{hardy05}. The precision of the measured quantities is not sufficient to add the decay of $^{26}$Si to the thirteen super-allowed transitions testing the electroweak Standard Model.

\section{Experimental procedure}

The experiment was performed at the Accelerator Laboratory of the University of Jyv\"{a}skyl\"{a}. We used the IGISOL technique
with the JYFLTRAP facility to separate pure samples of $^{26}$Si.

\subsection{Production and separation}

The ions were produced in light-ion induced fusion-evaporation reactions with a continuous 35~MeV proton beam having an average
intensity of 45 $\mu$A on a 2.3 mg/cm$^2$-thick $^{nat}$Al target. After being slowed down and thermalized in the gas cell of the
ion-guide~\cite{igisol}, the different recoil ions were accelerated to 30~keV. They were submitted to a mass separation in a
55$^{\circ}$ dipole magnet having a resolving power of 500, and the A$=$26 ions were injected into a buffer-gas filled RF-quadrupole
for cooling and bunching before injection into the first Penning trap of the JYFLTRAP tandem trap system ~\cite{jyfltrap1,jyfltrap2} for isobaric separation \cite{savard91}. The mass resolving power of the first trap was about 50,000 for this experiment and the cyclotron frequency set to select $^{26}$Si ions was 4134247~Hz.

The measurements were structured in cycles. A master cycle started with a 500~ms accumulation time in the RFQ followed by eight trap cycles and a decay measurement period. One trap cycle (0.231~s) was structured as follows: 100~ms (cooling) + 10~ms (dipole excitation) + 120~ms (mass selective quadrupole excitation). The ions were then ejected from the first trap, reflected by the second and recaptured again in the first one for the next trap cycle. As a consequence, the contaminants were removed because they could not pass the 2~mm diaphragm between the two traps and only $^{26}$Si returned to the first trap. This multiple injection method was favored in order to overcome the space charge limit of the purification trap. In parallel with one trap cycle, ions were cumulated into the RFQ for the next one. In the master cycle, the eighth trap cycle was followed by a final cleaning (0.231~s) of the accumulated bunches and by a 24.4~s decay measurement. The decay window was triggered by the trap extraction signal and during the decay measurement, the cyclotron beam was turned off in order to avoid any background in the experimental setup due to reactions on the target. Data were effectively taken over a period of 68 hours and we have accumulated a total of 3.559(2)~$\cdot$~10$^{6}$ $^{26}$Si ions.

\subsection{Experimental setup}

Purified samples of $^{26}$Si were implanted on a 0.5~inch wide movable tape placed at the end of the extraction beam line. The implantation spot was surrounded by an almost 4$\pi$ cylindrical plastic scintillator, 2~mm thick with a 12~mm entrance hole, used to detect the positrons emitted in the $\beta$$^+$ decay. The scintillation light was collected by two 2-inch photomultiplier tubes through a special light guide. The two photomultipliers were used in coincidence in order to remove most of the individual noise. The $\beta$$^+$-particle detection efficiency was about 90~\%~\cite{canchel05}. Three 60\% coaxial germanium detectors were placed around the plastic scintillator in the horizontal plane at $-$90$^{\circ}$ (Ge1), 0$^{\circ}$ (Ge3) and 90$^{\circ}$ (Ge2) angles with respect to the extraction beam line in order to provide $\beta$$-$$\gamma$ coincidence data. The detector labeled Ge3 was placed at 122.4~mm from the implantation point, whilst the other two were placed closer, at 29.3~mm (Ge1) and 30.4~mm (Ge2). The germanium crystals were surrounded by low-radioactivity lead bricks that reduced the $\gamma$ background by a factor of 4. The aim of the $\gamma$ detection was to measure the super-allowed BR and to monitor the background.

For the data taking we have used two independent data acquisition (DAQ) systems. The trigger for both DAQ systems was the coincident $\beta$ signal from the two photomultipliers and it was allowed only during the decay measurement time window of the master cycle. The first system, simple but fast, DAQ~A, was running in a cycle-by-cycle mode and had two predefined dead times $-$ 2 and 8~$\mu$s. The corresponding data will be referred to as Data1 and Data2. The time precision of this DAQ was determined by the clock of the PC on which it runs and it was far below 1 $\mu$s. The second system, DAQ~B, providing event-by-event data, had a predefined dead time of 100~$\mu$s and the corresponding data will be referred to as Data3. For the time stamp we used a 16-channel VME scaler that registered signals from a 1-MHz high precision clock generator. For both DAQ systems, the dead times were chosen to be longer that any possible event treatment by the electronics or data processing. The dead-time window was generated with a module having a precision better than 10~ns. DAQ~A registered only the time difference between the trap extraction signal and the subsequent event triggers. With the DAQ~B we could register also the energy signals from the germanium detectors in coincidence with the trigger signal.

\subsection{Search for $^{26}$Al$^m$ contamination}

As mentioned above, the Penning traps were used to provide a pure sample of $^{26}$Si on the tape for the life-time measurements. A possible contaminant was $^{26}$Al$^m$ having a half-life only three times longer than the one of $^{26}$Si. The contamination with $^{26}$Al$^m$ could come either from the reaction itself ($^{26}$Al$^m$ produced and selected together with $^{26}$Si), or from the decay of $^{26}$Si during the selection and transport to the detection system.

\begin{figure}[h]
\begin{center}
\includegraphics [width=0.45\textwidth] {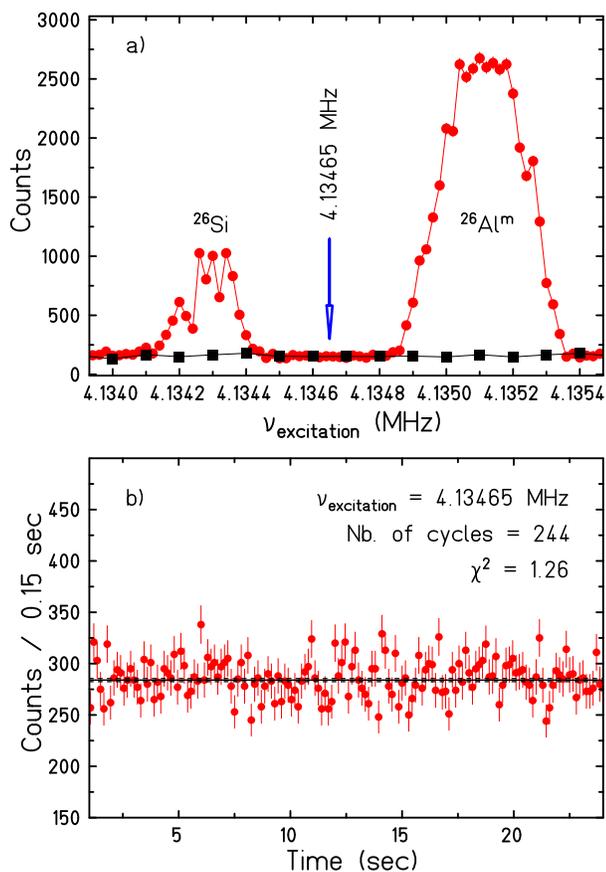}
\caption{a) Isobaric scan around $^{26}$Si and $^{26}$Al$^m$: circles $-$ 20~Hz scan and squares $-$ background scan. The background
was measured by inserting a beam stopper in the line. b) Decay spectrum when the excitation frequency is set between $^{26}$Si
and $^{26}$Al$^m$, as indicated in part a) of the figure.}
\label{fig:fig_qfreq_113}
\end{center}
\end{figure}

We have performed several tests in order to verify the purity of the samples. For the first test, the centering cyclotron frequency was switched off during the eighth cycle. Without centering, no ion is supposed to survive the extraction from the first trap after the last dipole magnetron excitation. This way, we could check that the magnetron excitation was strong enough to push all ions to radii bigger than 1~mm (the radius of the extraction hole) in the last trap cycle when we have the biggest ion cloud in the trap. The corresponding time spectrum accumulated during the decay time window is constant with a normalized $\chi^2$ of 1.

Another possible source of contamination with $^{26}$Al$^m$ could be an insufficient resolving power of the trap system. The 20~Hz step frequency scan presented in figure \ref{fig:fig_qfreq_113}a) was done in order to have a rough estimate of such a possible overlap. Then, in order to check the background measurement, we have fixed the excitation frequency to a value between the cyclotron frequencies for the selection of the two isobars. Using the same sequencing in the master cycle as for the half-life measurement, we have obtained the decay time spectrum presented in figure \ref{fig:fig_qfreq_113}b). The normalized $\chi^2$ indicated in fig.~\ref{fig:fig_qfreq_113}b) is obtained for a fit with a constant function. Using a degree-one polynomial for the fit gives a slope that is compatible with zero in the error bars.

\begin{figure}[h]
\begin{center}
\includegraphics [width=0.3\textwidth, angle=-90] {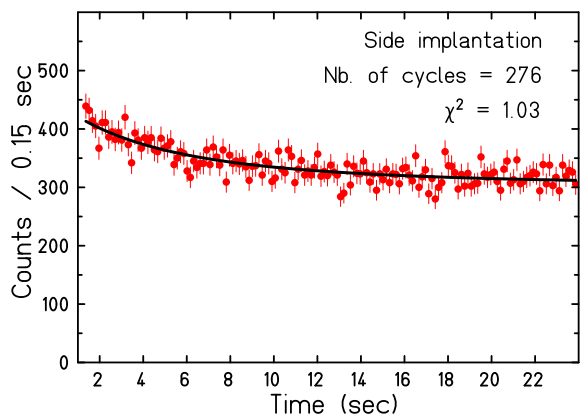}
\caption{Time spectrum for the test of side implantation. We present only the data points accumulated after the tape was moved
(see text for more details). The decaying component comes from the $^{26}$Si ions implanted on the entrance window of the plastic
scintillator which are not removed with the tape move.}
\label{fig:fig_r104_106}
\end{center}
\end{figure}

A final test to check for the initial conditions as far as the implanted sample was concerned was to verify that we implanted the ions entirely on the tape. To do so, we have changed only the decay measurement cycle as follows: after the extraction signal sent by the trap, we have measured the deposited activity for 1.3~s, moved the tape and continued to measure the activity until the end of the 24.4~s decay measurement window. If any activity was implanted somewhere else than on the tape, like \emph{e.g.} on the entrance window of the scintillator, the second part of the decay spectrum should still see the decay of $^{26}$Si and, subsequently, of $^{26}$Al$^m$. The resulting spectrum is presented in fig. \ref{fig:fig_r104_106} and one can easily see that such was the case. From this measurement we have deduced that 2.97(14)\% of the extracted $^{26}$Si was not implanted on the tape. This means that at the end of a master cycle, when the tape was moved, there was a remaining activity of $^{26}$Al$^m$ that had to be taken into account for the next cycle in the fitting function. As an example, for the highest counting rate per cycle during the experiment (about 550 implanted $^{26}$Si ions/cycle), one can estimate that a maximum of 2 atoms of $^{26}$Al$^m$ originating from the side implantation of the previous cycle were present at the beginning of the next cycle. We can also safely suppose that there is no $^{26}$Si left from one master cycle to the next. After this measurement, we added in the beam line a 50~mm thick collimator with a 10~mm diameter close to the scintillator entrance window in order to avoid the side implantation. The fit used for the runs after this change did not include anymore the influence of the side implantation.

\begin{figure}[!h]
\includegraphics [width=0.34\textwidth, angle=-90] {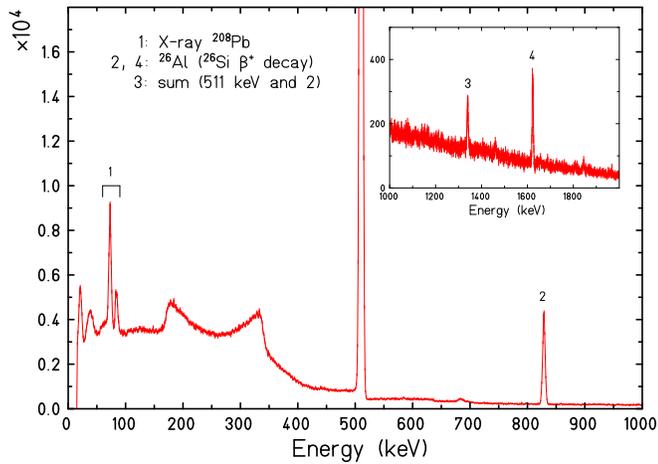}
\caption{Gamma spectrum in coincidence with a $\beta$ ray detected by the plastic scintillator. The X-rays are coming from the
low-radioactivity lead bricks used to reduce the $\gamma$ background.}
\label{fig:fig_gamma}
\end{figure}

As a further check for the absence of contaminants, we have analyzed the gamma-ray spectra registered in coincidence with the activity implanted on the tape. This allowed us to verify if there was any other $\gamma$-ray emitting contaminant in the implanted sample. The spectrum is presented in figure \ref{fig:fig_gamma}. The only $\gamma$-rays that are present come either from $\gamma$ or positron scattering in the lead bricks surrounding the germanium detectors, from the positron-electron annihilation, or from the $\beta$ decay of $^{26}$Si.

\section{Half-life results}

In this section, we will first discuss in detail the results from the three different data sets and the analysis procedure yielding the final half-life value with its statistical error. We will also discuss the influence of different parameters on this final result.

\subsection{Analysis procedure}

The fitting procedure can be found in more detail in \cite{blank04}. The first step of the analysis was a decay cycle selection.
We have selected all the cycles having a number of counts larger than 10. There were no significant changes in the life-time value when the minimum number of counts per cycle was varied up to 200. The accepted cycles were then corrected for the dead-time.

\begin{figure}[hh]
\begin{center}
\includegraphics [width=0.36\textwidth, angle=-90] {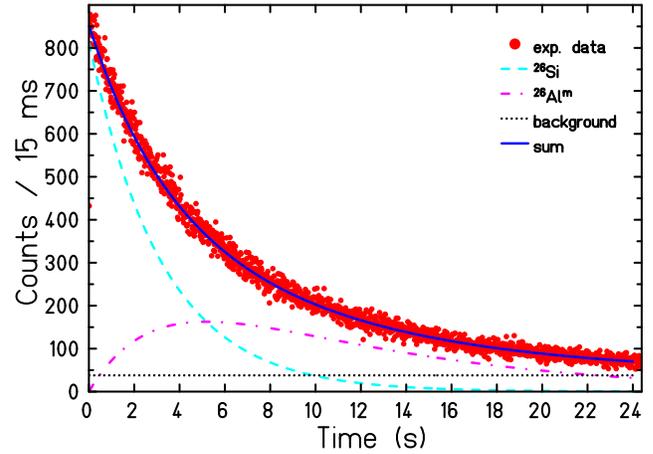}
\end{center}
\caption{The time distribution obtained for a single run (full circles). The full line is the result of the fit and the
contributions from $^{26}$Si, $^{26}$Al$^m$ and the background are represented separately. A half-life of 2.229(11)~s was
obtained for this run.}
\label{fig:fig_run98}
\end{figure}

\begin{figure*}[!ht]
\begin{center}
\includegraphics [width=0.57\textwidth, angle=-90] {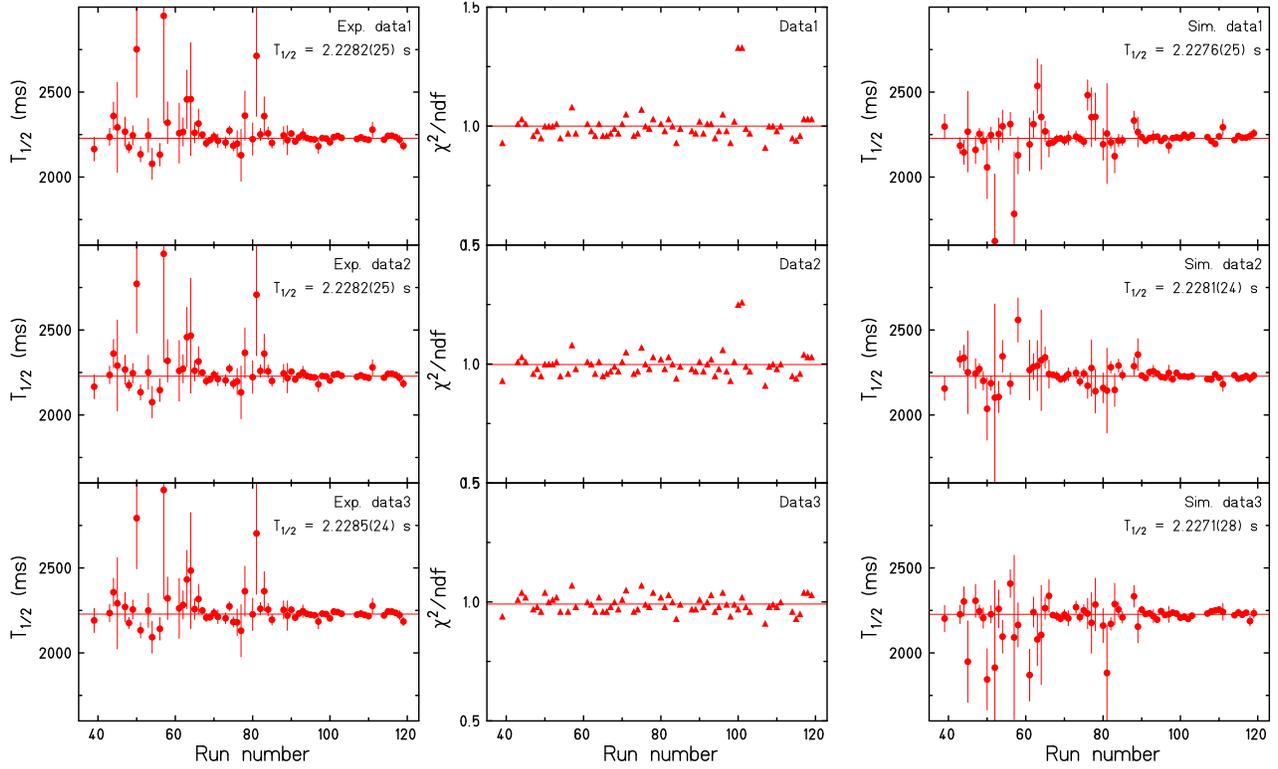}
\end{center}
\caption{Left: Experimental half-life as a function of run number for the three data sets. The error-weighted average value is 2.2283(25)~s (we cite here only the statistical error). Center: The normalized $\chi^2$ obtained from the fit of experimental data as a function of run number for the three data sets. Right: Half-life results from the simulated data as a function of run number for the three data sets. For the simulations, we have used a half-life value of 2.228~s for $^{26}$Si and the three data-sets were independently generated.}
\label{fig:fig_data_sim_chi2}
\end{figure*}

The next step was a cycle-by-cycle fit. The function used for the fit was defined to take into account the decay of $^{26}$Si and of its daughter, $^{26}$Al$^m$. Five parameters were used: the number of $^{26}$Si at the beginning of the decay cycle (N$_0^{Si}$), the half-life of $^{26}$Si (T$^{Si}_{1/2}$), a constant background, the half-life of $^{26}$Al$^m$ (T$^{Al}_{1/2}$ $=$ 6.3450(19)~s~\cite{hardy05}) and the correction factor that takes into account the side implanted $^{26}$Si ions. The last two parameters were fixed.

During the fit, we imposed the condition that the normalized $\chi^2$ has to be two or better in order to accept the cycle. This procedure rejects, \emph{e.g.}, cycles where problems with the HV of the RFQ occurred. Increasing the limit from 2 to 2000 for $\chi^2$ leaves the life-time unchanged. We have also excluded the first channel from the fit, corresponding  to the first 15~ms of the decay cycle that includes the period when we could still have incoming ions from the trap. This decision was supported by the fact that the results including the first channel were quite different (up to 0.7\%) from the ones excluding it. We varied the number of excluded channels at the beginning of the time spectra from 2 to 30 but with no significant change (less than 0.04\%) appearing in the resultant life-time.

All in all, about 2 to 3\% of the cycles were rejected because the fit did not converge or the $\chi^2$ was higher than 2. The accepted cycles were further grouped into runs and the cumulated decay spectra were fitted again run-by-run with the same procedure. One run contained between several to 400 cycles. Figure~\ref{fig:fig_run98} shows the experimental decay-time spectrum decomposed into its different contributions from the decay of $^{26}$Si, $^{26}$Al$^m$, and of the background for one run.

The fit results of the three data sets and the associated normalized $\chi^2$ as a function of the run number are presented in figure \ref{fig:fig_data_sim_chi2}, left and center. The important scattering and the associated error bars for the half-life values from run 39 to 93 are due to a low production/selection efficiency for $^{26}$Si. We obtain a mean half-life of 2.2282(25)~s for Data1, 2.2282(25)~s for Data2 and 2.2286(24)~s for Data3. The resulting experimental half-life for the $^{26}$Si ground state is 2.2283(25)~s. This value is the weighted mean of the three data sets and the statistical error is chosen to be the biggest one since the data sets are not independent measurements.

In parallel, for each selected cycle, we have generated simulated data for which all characteristics except the half-life were determined by the fit of the corresponding experimental cycle. We used a half-life of 2.228~s for the generation of the simulated spectra. The simulated data were then analyzed with the same procedure as the experimental data. The results obtained for the simulated data are summarized on the right side of figure \ref{fig:fig_data_sim_chi2}.

\subsection{Error budget}

The experimental half-life value cited above includes only the statistical error obtained from the fit of time spectra of the 3 data sets. In the following, we shall discuss other sources of errors for the measured value like, \emph{e.g.}, fixed parameters in the fit function, systematic errors due to experimental conditions, etc.

\subsubsection{Systematic errors associated with the fitting procedure}

As previously mentioned, we used a five parameter function to fit the experimental spectra. Two of these parameters were fixed: the life-time of $^{26}$Al$^m$ and the percentage of side implanted $^{26}$Si. In order to take into account the errors on these parameters, we have included them in the final result for the half-life of $^{26}$Si by changing the fixed parameter values within one sigma. This gives an error of 0.3~ms that will be referred to as the systematic error due to fixed parameters (SEFP).

Also, the half-life results from the three different data sets are slightly different from each other. To take this effect into account, we have calculated the sum of squared differences between each value and the central mean value. This gives a systematical error of 0.2~ms to which we shall refer to as the error due to dead-time corrections (SDT).

\subsubsection{Experimental conditions and systematic errors}

During the experiment we have made several modifications to the electronics setup in order to check for systematic errors.

\begin{table}[h]
\begin{center}
\caption{Error budget for the $^{26}$Si life-time measurement. SHV-CFD is the error due to detector bias and threshold of constant fraction discriminators, SEFP is the error due to fixed parameters in the fit and SDT is due to dead-time corrections. The individual values are added quadratically to calculate the final error on the half-life of $^{26}$Si.}
\begin{tabular}{l r}
  \hline\hline
  Source  & Uncertainty (ms) \\
 \hline
   Statistical error &  2.5 \\
   SHV-CFD &  1.0 \\
   SEFP & 0.3  \\
   SDT & 0.2  \\
              \\
   Final error & 2.7  \\
  \hline
\label{errortab}
\end{tabular}
\end{center}
\end{table}

The HV of the photomultipliers was changed during the experiment from -1.73~kV to -1.92~kV along with the thresholds of the constant fraction modules used to trigger the photomultiplier signals. The two photomultipliers were always biased at the same value using one HV module (Ortec HV-556) with two identical outputs. The experimental data presented in figure \ref{fig:fig_data_sim_chi2} can be structured in three main groups with respect to the HV and constant fraction threshold values: runs 39-93, runs 96-99 and runs 103-119. Results from the fits of either group are consistent with each other at one sigma. Nevertheless, they introduce a systematical error (referred to as SHV-CFD) calculated as the sum of squared differences between each value and the half-life mean value weighted by the respective errors that gives a systematical error of 1~ms.

In table \ref{errortab} we quote the contributions from different sources to the final error on the experimental half-life value.

\subsection{Final experimental result for the half-life}

\begin{figure}[h]
\begin{center}
\includegraphics [width=0.33\textwidth, angle=-90] {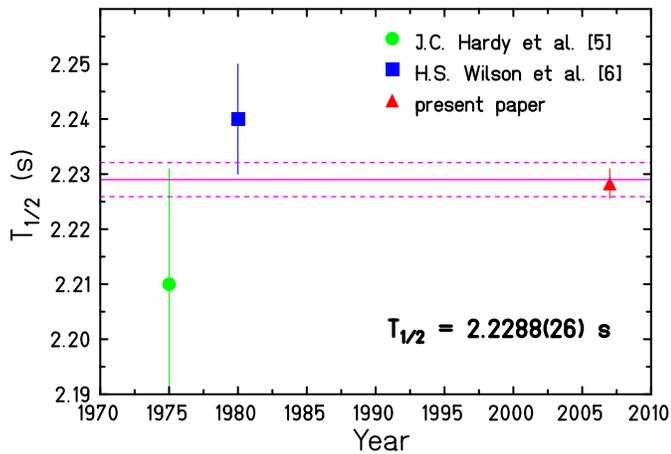}
\end{center}
\caption{Comparison of the present measurement of the $^{26}$Si half-life with previous measurements reported in the literature.
The full line indicates the weighted mean of all the existing measurements; the dashed lines indicate the error on the mean value.}
\label{fig:fig_t12}
\end{figure}

The final result for the half-life of the $^{26}$Si ground state is 2.2283(27)~s. Previous measurements of the ground state half-life were reported by Hardy \cite{hardy75} (2.210(21)~s) and Wilson \cite{wilson80} (2.240(10)~s). In figure \ref{fig:fig_t12}, one can see that the agreement is very good between these values and our measurement.

\section{Branching ratio for the 0$^+$ $\rightarrow$ 0$^+$ transition}

As previously mentioned, the BR for the super-allowed decay was already measured with a precision of about 1\% \cite{hardy05}. It was obtained by measuring the absolute non-analogue $\beta^+$ branching for the most intense $\gamma$ transition (829~keV de-exciting the E$_x$=1058~keV energy level in $^{26}$Al) and the relative intensities of the other $\gamma$ transitions relative to the 829~keV transition. This accuracy is not enough if we want to know the $\mathcal{F}t$ value for $^{26}$Si with a precision of several 10$^{-4}$.

The main purpose of the present experiment being the half-life measurement, we were not aiming to achieve the required precision on the BR. Nevertheless, we have analyzed the $\gamma$ spectra of the three germanium detectors and we will present in the following the procedure we used to determine the BR.

The total photopeak $\gamma$ efficiency ($\varepsilon$$_T$) of the germanium setup was measured using standard calibration sources of $^{134}$Cs, $^{137}$Cs, $^{60}$Co, $^{133}$Ba and $^{228}$Th. The $^{60}$Co source had an activity known with a precision better than 0.1\%. The first step in the analysis was the direct determination of the efficiency curve from the source measurements. This efficiency curve had then to be corrected for $\beta$-$\gamma$ or $\gamma$-$\gamma$ summing effects. These corrections can be derived from simulations and one has to take into account as exhaustively as possible all the mechanisms by which a $\beta$ or a $\gamma$-ray can produce charges in the germanium crystals. For example, in the case of $^{137}$Cs, the correction should be close to 1 as there is only one $\gamma$-ray and no $\beta$-$\gamma$ summation (Q$_{\beta^-}$ being too low to have electrons with a kinetic energy high enough to arrive in the germanium crystals).

We have used the GEANT4 package \cite{geant4} to calculate the correction factors for the efficiency curve. The experimental setup defined in the simulations included the vacuum chamber, the lead used to screen the germanium detectors from the background radioactivity and the germanium detectors. The calibration sources were defined as being point-like since there is no significant change in the correction factors if one uses finite size sources and we took into account their complete decay scheme. We have started with the simulation of single $\gamma$-rays (thus, no summing effects) generated from the source position and counted the number of events in the photopeak (N$_{single}^{\gamma}$). The next step was the simulation of the complete decay scheme of each source so that the $\beta$-$\gamma$ or $\gamma$-$\gamma$ summing effects could be taken into account. Then, the number of events in each photopeak divided by N$_{single}^{\gamma}$ for the same energy was the correction factor used to obtain the corrected experimental single gamma efficiency curve.

\begin{table}[h]
\begin{center}
\caption{ Correction factors obtained from simulations for individual $\gamma$ rays of calibration sources for each germanium detector. }
\begin{tabular}{l r r r r}
  \hline\hline\rule{0pt}{1.3em}
  Source & E$_\gamma$ (keV) & Ge1 & Ge2 & Ge3 \\[1ex]
  \hline \rule{0pt}{1.3em}
  $^{60}$Co & 1173 & 0.951(10) & 0.951(10) & 0.998(3) \\
            & 1332 & 0.960(8) & 0.960(8) & 0.998(3) \\[1ex]
  $^{133}$Ba & 276 & 0.811(38) & 0.814(37) & 0.947(12) \\
             & 302 & 0.876(25) & 0.879(24) & 0.985(5) \\
             & 356 & 0.885(23) & 0.894(21) & 0.953(10) \\[1ex]
  $^{134}$Cs & 569 & 0.883(24) & 0.885(23) & 0.983(6) \\
             & 604 & 0.924(15) & 0.929(14) & 0.990(3) \\
             & 795 & 0.929(14) & 0.929(14) & 0.989(4) \\[1ex]
  $^{137}$Cs & 661 & 0.996(1) & 0.996(1) & 0.995(3) \\[1ex]
  $^{228}$Th & 2614 & 0.913(18) & 0.913(18) & 0.989(5) \\
  \hline
\label{coeftab}
\end{tabular}
\end{center}
\end{table}

\begin{table*}[ht]
\begin{center}
\caption{ The absolute $\beta^+$ BR for the most intense $\gamma$-line, 829~keV (BR(1058~keV)) and the relative intensity of the 1622~keV transition with respect to the 829~keV line ($\gamma_{1622}$/$\gamma_{829}$) are reported for each germanium detector and compared with the adopted values in \cite{hardy05}. The mean values obtained after averaging over the results of the three detectors are also compared with the adopted values in \cite{hardy05}.}
\begin{tabular}{l r r r r r}
  \hline\hline\rule{0pt}{1.3em}
    & Ge1 & Ge2 & Ge3 & Mean values & \cite{hardy05} \\[1ex]
  \hline \rule{0pt}{1.3em}
  BR(1058~keV) (\%) & 21.03(94) & 20.15(73) & 22.19(67) & 21.21(64) & 21.8(8)  \\[1ex]
  $\gamma_{1622}$/$\gamma_{829}$ &  &  &  & 0.1301(62) & 0.1265(36)  \\[1ex]
  BR(0$^+$ $\rightarrow$ 0$^+$) (\%) &  &  &  & 75.69(232) & 75.09(92)  \\[1ex]
  \hline \rule{0pt}{1.3em}
\end{tabular}
\label{BRtab}
\end{center}
\end{table*}

\begin{figure}[h]
\begin{center}
\includegraphics [width=0.33\textwidth, angle=-90] {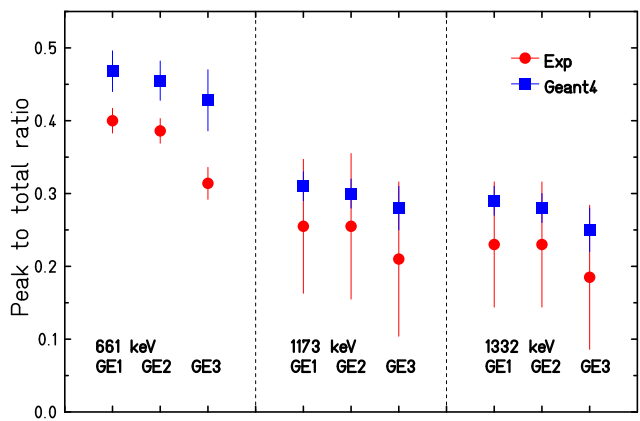}
\end{center}
\caption{Comparison between the calculated (squares) and the experimental (circles) peak-to-total (P/T) ratios for the $^{137}$Cs, $^{60}$Co. The experimental P/T ratios for the $^{60}$Co source were determined by coincidence between a pair of germanium detectors.}
\label{fig:figPT}
\end{figure}

We have also compared the calculated peak-to-total (P/T) ratios for the $^{137}$Cs and $^{60}$Co sources with the experimental ones. The results are represented in figure \ref{fig:figPT}. One can easily see that there is a systematical difference between the experiment and the calculations of about 20\%. This can come from a lack of knowledge about the materials surrounding the implantation site that are very important for the Compton scattering. We decided to take this difference into account by adding quadratically 20\% of (1 $-$ correction factor value) to the previous errors on the correction factors.

A summary of the correction factors obtained for the sources used for $\gamma$ calibration for each germanium detector is given in table \ref{coeftab}. From the corrected efficiency curve we have then determined the single gamma photopeak efficiency for the 829~keV and 1622~keV transitions in the decay of $^{26}$Si.

To calculate the correction factors to be applied to the experimental number of events for each of the two transitions, we have also simulated the $^{26}$Si source taking into account the finite source size and the $\gamma$-branching ratios as given in the literature \cite{hardy05}. We have also taken into account the positrons emission in the $\beta$$^+$ decay and their annihilation in the materials surrounding the experimental setup. This is important because the 511~keV $\gamma$-ray plays an important role in the summing effects for the germanium spectra. The same procedure as for the calibration sources was then applied in order to determine the factors needed to correct for summing effects.

Using the single gamma efficiency, the corrected number of events in the photopeak and the number of implanted $^{26}$Si obtained from the fit of the decay time curve, we could then determine the absolute intensity of the 829~keV transition (BR(1058~keV)) for each of the three germanium detectors, and the relative intensity of the 1622~keV transition with respect to the 829~keV line ($\gamma_{1622}$/$\gamma_{829}$) averaged over the three detectors. The results are presented in table \ref{BRtab} and compared with the adopted values in \cite{hardy05}. We deduce then an absolute $\beta$-decay branch for non-analogue transitions of 24.31(232)\% resulting in a absolute $\beta$-decay branch for the super-allowed transition of 75.69(232)\%. For the transitions that were not observed in our experiment we have used the relative intensities from \cite{hardy05}.

\section{Conclusions}

We have performed a high-precision measurement of the half-life of $^{26}$Si. The half-life was determined by detecting the $\beta$ particles from the decay of a $^{26}$Si source produced and separated at the Accelerator Laboratory of the University of Jyv\"{a}skyl\"{a} using the IGISOL technique with the JYFLTRAP facility. The result of T$_{1/2}$~= 2.2283(27)~s obtained in this work is in agreement with older half-life values from the literature. The present result is a factor of 4 more precise than the previous measurements. The error-weighted mean value from all reported measurements is 2.2288(26)~s. With this precision of 14 parts in 10$^{4}$, the half-life of $^{26}$Si is precise enough to contribute to the test of the CVC hypothesis.

We have also measured the BR value for the super-allowed transition and obtained a value of 75.69(232)\% that has a similar precision as previous measurements \cite{hardy05}. Averaging over the presently known super-allowed BR we obtain a value of 75.17(86)\%. Using the new values for the correction factors $-$ $\delta'_R$, $\delta_C$, $\delta_{NS}$ $-$ and the statistical rate function, f, as given in \cite{hardy05,towner07} the average value of $\mathcal{F}t$ for $^{26}$Si becomes 3060(37)~s.

In order to include $^{26}$Si in the high precision measurements of super-allowed $\beta$ decays, one needs to improve the precision of Q$_{EC}$ and the super-allowed BR. The Q$_{EC}$ has already been remeasured at JYFLTRAP and the results will be published in the near future. It remains then to improve the precision on the BR value which is one of our future priorities.

\begin{acknowledgement}

The authors would like to acknowledge the continuous effort of the whole Jyv\"askyl\"a accelerator laboratory staff for ensuring
a smooth running of the experiment. We are grateful to our colleagues from the laboratory LNE-LNHB at CEA Saclay for the fabrication and calibration of the $^{60}$Co source. This work was supported in part by the Conseil R\'egional d'Aquitaine and by the European
Union 6th Framework Programme "Integrated Infrastructure Initiative - Transnational Access", Contract No. 506065 (EURONS).
We also acknowledge support from the Academy of Finland under the Finnish Center of Excellence Programme 2006-2011 (Project No.
213503, Nuclear and Accelerator Based Physics Programme at JYFL).

\end{acknowledgement}

\end{document}